\documentclass[sigconf]{acmart}

\usepackage{booktabs}
\usepackage{multirow}
\usepackage{array, makecell} %
\usepackage{comment}
\usepackage[utf8]{inputenc}
\usepackage{cancel}

\newcommand{\da}[1]{\textcolor{red}{#1}}
\newcommand{\vn}[1]{\textcolor{black}{#1}}

\definecolor{lightblue}{RGB}{100, 150, 255}

\acmYear{2026}\copyrightyear{2026}
\setcopyright{cc}
\setcctype[4.0]{by}
\acmConference[ ]{ }{ }{ }
\acmBooktitle{ }
\acmDOI{10.1145/3803437.3805253}
\acmISBN{979-8-4007-2636-1/26/07}

\setcopyright{none}
\renewcommand\footnotetextcopyrightpermission[1]{}

\begin{document}

\title{How Analysts Use AI in High-Stakes Crime Linkage: An Industrial Study}


\author{Jessica Woodhams}
\affiliation{%
  \institution{University of Birmingham, UK}
  \city{}
  \country{}
}
\email{j.woodhams@bham.ac.uk}

\author{Amy Burrell}
\affiliation{%
  \institution{University of Birmingham, UK}
  \city{}
  \country{}
}
\email{a.burrell@bham.ac.uk}

\author{Wanyin Li}
\affiliation{%
  \institution{University of Reading, UK}
  \city{}
  \country{}
}
\email{wanyin.li@reading.ac.uk}

\author{Fahim Ahmed}
\affiliation{%
  \institution{Imperial College London, UK}
  \city{}
  \country{}
}
\email{fahim.ahmed22@imperial.ac.uk}

\author{Matthew Tonkin}
\affiliation{%
  \institution{University of Leicester, UK}
  \city{}
  \country{}
}
\email{mjt46@leicester.ac.uk}

\author{Jan Lemeire}
\affiliation{%
  \institution{Vrije Universiteit, Belgium}
  \city{}
  \country{}
}
\email{jan.lemeire@vub.be}

\author{Arkady Konovalov}
\affiliation{%
  \institution{University of Birmingham, UK}
  \city{}
  \country{}
}
\email{a.konovalov@bham.ac.uk}

\author{Steven Frisson}
\affiliation{%
  \institution{University of Birmingham, UK}
  \city{}
  \country{}
}
\email{s.frisson@bham.ac.uk}

\author{Mark Webb}
\affiliation{%
  \institution{National Crime Agency, UK}
  \city{}
  \country{}
}
\email{mark.webb@nca.gov.uk}

\author{Sarah Galambos}
\affiliation{%
  \institution{National Crime Agency, UK}
  \city{}
  \country{}
}
\email{sarah.galambos@nca.gov.uk}

\author{Vesna Nowack}
\affiliation{%
  \institution{Imperial College London, UK}
  \city{}
  \country{}
}
\email{v.nowack@imperial.ac.uk}

\author{Dalal Alrajeh}
\affiliation{%
  \institution{Imperial College London, UK}
  \city{}
  \country{}
}
\email{dalal.alrajeh04@imperial.ac.uk}



%
%

\renewcommand{\shortauthors}{Woodhams et al.}

\begin{abstract}
Crime linkage analysis is used in many countries to identify series of offences that may have been committed by the same individual. In practice, specialist analysts manually search for behavioural and situational connections across large crime databases, an effort that is time-consuming, cognitively demanding, and can involve repeated exposure to disturbing material. To support this work,  an Artificial Intelligence (AI)-enabled decision-support tool was co-developed with a UK law enforcement agency to assist analysts in identifying likely crime linkages.

This paper reports an industrial evaluation of the crime-linkage tool. We conducted a mixed-methods usability study combining direct observation, eye-tracking, mouse-tracking, and surveys to examine how analysts engage with AI predictions and with the model features presented as explanations. Our findings show that analysts used the AI predictions selectively and frequently validated them against behavioural (non-AI) evidence, reflecting partial trust and an ongoing reliance on established analytical practices. We also found that analysts attended to the presented model features and valued their availability, while identifying opportunities to improve how explanations are presented and integrated into the workflow. Overall, our results highlight the need for AI-enabled decision-support tools to better integrate explanations and traditional analytical methods, and demonstrate the importance of in-situ evaluation for engineering usable and trustworthy AI in high-stakes settings.

\end{abstract}


\maketitle


\section{Introduction}
\label{sec:introduction}




Serial crimes, in which an offender commits multiple offences against different victims over time, pose particular challenges for investigators attempting to identify and connect linked incidents. One major challenge is that many criminals do not leave physical forensic evidence at crime scenes, making it hard to later link crimes and establish a series~\cite{Amankwaa2021effectiveness}. Specialist units of highly trained expert analysts exist in different countries of the world who, instead, use behavioural information about the commission of the crime to identify potential linked series. This practice is called crime linkage. 

Over time, databases of crime-related data grow to hold the behavioural details of tens of thousands of crimes. For example, the UK's database for a stranger sexual offences contains data related to around 37,000 cases, including information about where and when offences happened, as well as the behaviour demonstrated by the offender during the crime~\cite{Tonkin2025CrimeLinkage}.

Analysts, therefore, have the challenging task of searching within these databases for crimes likely committed by the same individual. Often their task involves comparing a single offence (called \textit{the index offence}) to other crimes in the database. They look for other crimes that not only share similar behaviour but this shared behaviour must also be distinctive in some way from the behaviour of other offenders in the database. These two conditions are referred to as the assumptions of behavioural consistency and distinctiveness~\cite{Woodhams2019Linking} and have been the subject of academic research for two decades. The research generally supports the view that criminals are sufficiently consistent and distinctive in their behaviour for their crimes to be linked behaviourally~\cite{Bennell2014Linking,Santtila2005Behavioral,Yokota2007Application}.

%

Qualitative research with crime linkage analysts has shown that current tools for interrogating crime databases are often inadequate, with analysts expressing concern that important potential links may be missed~\cite{Burrell2011Preliminary}. This has motivated sustained research into methods that can assist the crime linkage process. In recent years, scholars have explored statistical and machine-learning (ML) approaches that aim to semi-automate parts of the linkage task by predicting which crimes are most likely to be connected~\cite{Tonkin2017Using,TonkinWeeks2021,Tonkin2025CrimeLinkage}. These models are intended to underpin decision-support systems that surface empirically informed predictions to analysts.

However, the usefulness of such systems depends not only on the accuracy of the underlying models, but also on how their predictions are communicated to end users~\cite{Ribeiro2016}. \vn{This challenge spans both the design of user-facing explanations and the broader system engineering process~\cite{alrajeh2026datadependentgoalmodelingmlenabled}. In this paper, we focus on the former by examining how crime-linkage analysts engage with AI-generated predictions and explanations in situ, and how these outputs are integrated with (and validated against) behavioural evidence and established analytical practices.}

Prior work in human-AI interaction has shown that the presentation of ML outputs influences interpretability, trust, and operational acceptance of AI-enabled tools~\cite{Amershi2019,Lipton2018}. \vn{Studies in the law enforcement domain similarly highlight the importance of providing explanations for AI-generated predictions and decisions~\cite{Herrewijnen2024Requirements,Nowack2025UserCentredAI}, as explainability directly supports trustworthiness, a key quality requirement for AI-enabled decision-making systems in this context~\cite{Nowack2025UserCentredAI}. While prior work has examined transparency and explainability of AI tools in policing~\cite{Hepenstal2025Towards}, further research is needed to understand the practical usability of AI-enabled decision-support tools in high-stakes operational settings.}

In this paper, we report the results of a mixed-method usability evaluation of an AI-enabled crime-linkage decision-support tool deployed within a UK law enforcement agency \vn{called the National Crime Agency (NCA), the lead agency against organised crime, human and drug trafficking, weapons and cybercrime\footnote{https://www.nationalcrimeagency.gov.uk}.} The decision-support tool, \vn{LATIS (formerly known as DST)}~\cite{Tonkin2025CrimeLinkage}, was developed by a multidisciplinary research team for the NCA's Serious Crime Analysis Section (SCAS), the UK's unit responsible for behavioural linking of serious sexual offences~\cite{NCASeriousCrimeAnalysis}. The tool is built on ML prediction models trained on real crime data and returns, from the  database, crimes most likely to be linked to a user-specified index offence. Following extensive testing of predictive performance, a visual interface was developed to present results to analysts and support their decision making by combining AI predictions with behavioural (non-AI) evidence. The aim of this paper was to address the following research questions:



\noindent \textbf{RQ1}:
What experiences and perceptions do analysts report when using an AI-enabled crime-linkage tool?

\noindent \textbf{RQ2}:
Do analysts attend to all the features \vn{underlying AI predictions when interacting with the tool?}

\noindent \textbf{RQ3}:
To what extent do analysts integrate their traditional \vn{analytical} methods to complement AI predictions?



\vspace{3mm}
\noindent The contributions of this paper are as follows:

\vn{\noindent\textbf{First industrial evaluation of an AI-enabled crime-linkage tool using mixed methods:}
We present, to the best of our knowledge, the first industrial evaluation of an AI-enabled crime-linkage tool, combining a mixed-methods usability study with operationally realistic tasks performed by analysts in a law enforcement agency using real crime series.}

\vn{\noindent \textbf{Evidence on usability, perceived value, and adoption considerations in operational practice:}
We provide empirical evidence of analysts' engagement with the tool, including overall positive usability feedback (notably ease of use), alongside more mixed perceptions of usefulness, highlighting concrete barriers and opportunities for improvement before wider deployment.}

\vn{\noindent \textbf{Empirical insights into interaction with AI predictions and feature-based explanations:}
We characterise how analysts attend to and interpret AI outputs and the underlying model features, supported by qualitative feedback and eye-tracking results showing attention across features,
demonstrating the value of tool's transparency and the alignment of the explanation design with analysts' natural decision priorities.}

\vn{\noindent \textbf{Design implications for integrating AI support with traditional analytical methods:}
We identify that analysts frequently cross-check AI suggestions using traditional (non-AI) behavioural analysis (e.g., the behavioural matrix), supported by interaction evidence, and derive actionable design implications for integrating such complementary analytical methods directly into AI-enabled tools to improve trust, efficiency, and operational acceptability.}

The rest of the paper is structured as follows. The next section provides the background. We describe our study design in Section~\ref{sec:evaluation} and study findings in Section~\ref{sec:results}. Ethical considerations can be found in Section~\ref{section:ethics}, threats to validity in Section~\ref{sec:threats} and related work in Section~\ref{sec:related}. We conclude in Section~\ref{sec:conclusion}.


\section{Background}
\label{sec:background}

\subsection{Crime linkage process} 
Crime linkage is a type of behavioural analysis that aims to determine whether crimes were committed by the same offender, by examining similarities in behaviour across the offence of interest (\textit{the index offence}) and other crimes~\cite{WoodhamsTonkin2017}. Research has  demonstrated that offenders are both sufficiently consistent and distinctive from one another in their offending behaviour that their crimes can be linked as a series~\cite{Bennell2014Linking,Santtila2005Behavioral,Yokota2007Application}.

The first stage in crime linkage involves an analyst familiarising themselves with the offender behaviour seen in the index offence~\cite{Woodhams2007Case}. Research of crime series has shown that offenders display consistency in their behaviour between their crimes and that crimes from the same series also tend to occur in geographical and temporal proximity to one another~\cite{Bennell2014Linking,Woodhams2021Descriptive}. Therefore, as well as similarity in patterned behaviour exhibited by an offender (commonly known as modus operandi, or MO), analysts also use measures of proximity to make crime linkage decisions~\cite{Burrell2011Preliminary}. This can include characteristics of the crime context, such as victim age, location(s) chosen for the crime, and the time and day of the crime~\cite{Hazelwood2004Linkage,Labuschagne2006Use}.

Having familiarised themselves with the crime and offender actions, the analyst then makes a list of salient behaviours that they think the suspect may repeat in other crimes. These salient behaviours guide the analyst when they search police databases for other crimes that are behaviourally similar to the index offence and also distinctive in behaviour from other offences of the same type. Where such crimes are found in databases, the analyst familiarises themselves with the details of these crimes too, compares and contrasts crimes with the index offence to decide if they are potentially linked (and committed by the same offender)~\cite{Woodhams2007Case,Burrell2011Preliminary,Davies2018Practice}.

\subsection{Development of the AI-enabled crime-linkage decision-support tool}
\vn{As part of our earlier work, we co-developed with NCA} an AI-enabled tool that assists analysts in identifying offences linked to the index offence. \vn{Prior to the tool development, the researchers conducted focus groups to gain a deeper understanding of crime-linkage processes and analyst needs. During these sessions, analysts expressed the need for (i) an automated tool to support their decision-making, (ii) transparency in how the tool's predictions are generated, and (iii) retaining responsibility as the final decision-makers.}

The tool was then designed through workshops with analysts and leveraged ML models trained on behavioural, temporal, and geographical features of offences~\cite{Tonkin2025CrimeLinkage}. Its aim is to enable analysts to explore potential linkages, inspect the behavioural evidence underlying AI predictions, and verify findings against non-AI data sources. In doing so, it would reduce manual workload and limit analysts' exposure to disturbing material.

\begin{figure*}[t]
\centering
\includegraphics[width=\textwidth]{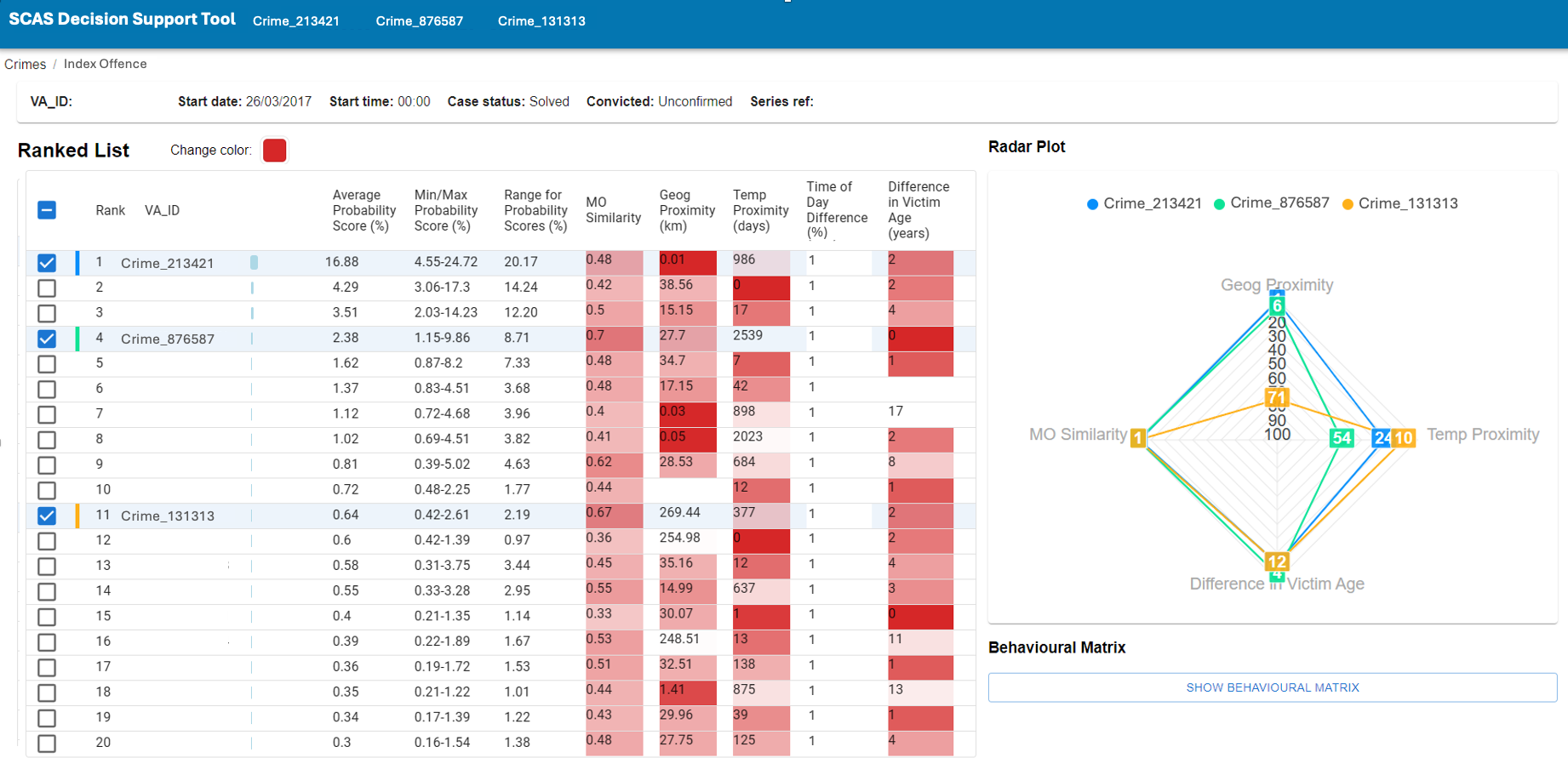}
\caption{Prototype showing crime linkage predictions, underlying model features and behavioural similarities between the index offence and three other crimes. Identifiers are removed for anonymity.}
\label{figure:screenshot}
\end{figure*}

Follow-up workshops indicated the need of a visualisation that displays (i) ranked search results indicating the likelihood that an offence is linked to the index case, and (ii) transparent explanations, such as the features contributing to each ranking. The analysts also requested mechanisms to inspect behavioural similarities across crimes, consistent with current practice using behavioural matrices. As a result, the final interface design combined a ranked-list view displaying linkage probabilities with feature-level explanations and optional comparison views for selected offences, presented as a radar plot for feature-level comparison and a behavioural matrix highlighting shared behaviours (Figure~\ref{figure:screenshot}). This design was validated by the SCAS senior leadership.

\noindent
\textbf{Ranked list display.} 
The ranked list presents crimes ordered by their predicted probability of linkage to the index offence (values between 0 and 1). Each entry combines a numerical score with a horizontal bar to provide an at-a-glance comparison of linkage strength, similar to the mini-map feature in \textit{Podium}~\cite{wall2017podium}. Five key AI features (MO similarity, geographical proximity, temporal proximity, time of day difference and difference in victim age) are visualised using coloured cells, with saturation indicating each feature's relative contribution to the overall score. Displaying the contributions of individual features enhances transparency and allows analysts to identify not only high-probability cases but also the features driving those predictions.

\noindent
\textbf{Radar plot display.} 
Radar plots support detailed comparison between the index offence and a selected subset of crimes. Each spoke represents one of five overarching features used in the model. Feature values are normalised to percentile scores to reflect their distinctiveness relative to the full dataset. To ensure interpretability, distance-based measures (e.g., spatial, temporal, victim age) are reverse-scaled so that higher values consistently indicate greater similarity. Multiple crime-index pairings are overlaid in a single plot using colour coding, and precise values are displayed via hover interaction. 

\begin{figure}[t]
\centering
\includegraphics[width=0.9\columnwidth]{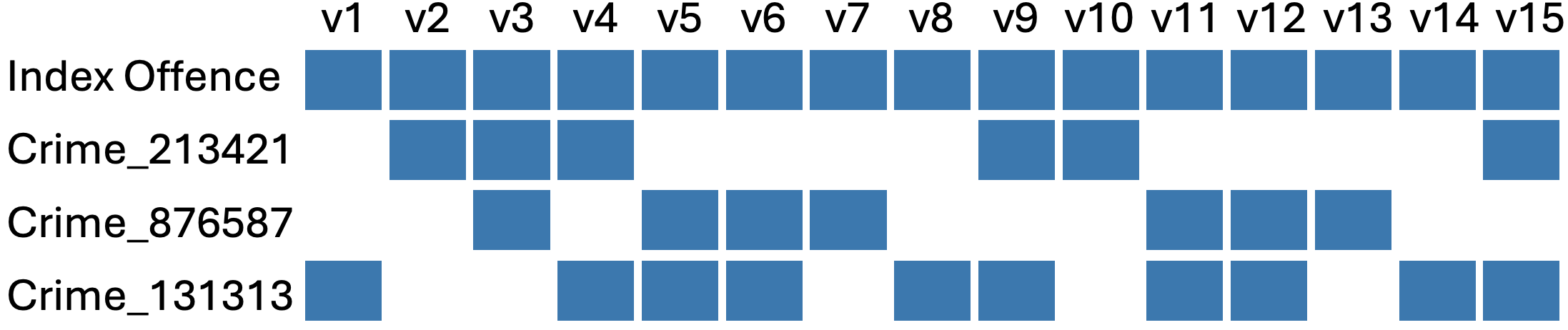}
\caption{Behavioural matrix showing matching variables between the index offence and comparison crimes (coloured cells). Actual variable names are omitted to preserve confidentiality.}
\label{figure:behav_matrix}
\end{figure}

\noindent
\textbf{Behavioural matrix.} 
For fine-grained behavioural analysis \vn{(independent of AI predictions)}, a matrix view automatically populates behavioural variables present in the index offence and selected comparison crimes (Figure~\ref{figure:behav_matrix}). Columns correspond to behavioural variables, and rows to the index and comparison offences. Coloured cells indicate the presence of a behaviour, while blank cells indicate absence. This design, inspired by scatterplot tables and colour maps~\cite{sedrakyan2019guiding,blumenschein2018smartexplore}, mirrors existing manual analytic practices~\cite{craik1994linking} while reducing the time required to create such matrices. Due to the large number of variables \vn{(typically several dozen and sometimes over one hundred)}, the matrix is implemented as a pop-out window from the main display.

The tool design provided a transparent and interactive interface to support analysts in reviewing generated predictions, exploring feature-level explanations, and examining behavioural evidence, and addressed usability and interpretability requirements identified during earlier workshops. 

\section{Study Design for the Evaluation of the AI-enabled crime linkage tool}
\label{sec:evaluation}

\begin{figure}[t]
\centering
\includegraphics[width=0.85\columnwidth]{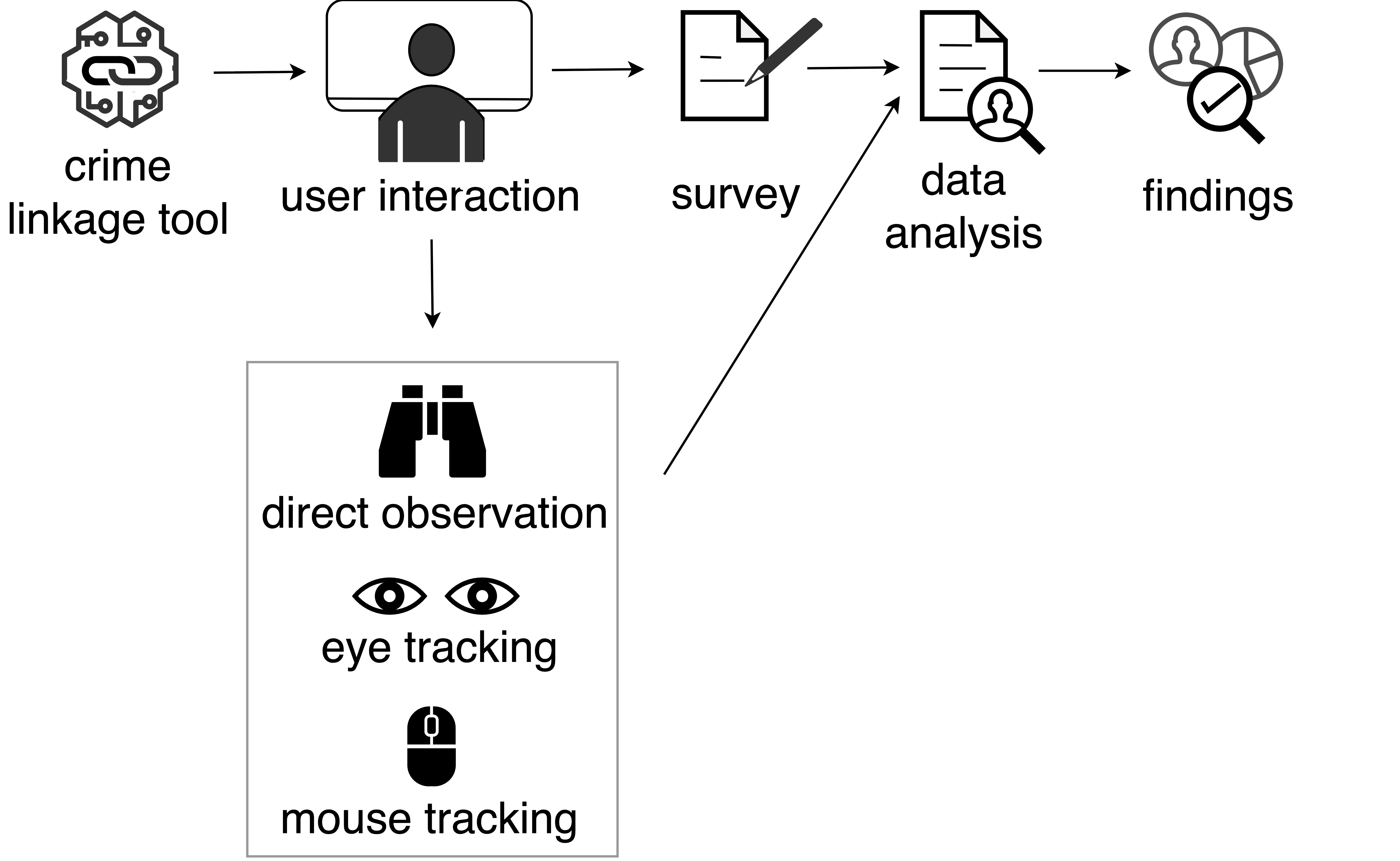}
\caption{\vn{Overview of the study design used to evaluate the AI-enabled crime linkage tool.}}
\label{figure:evaluation}
\end{figure}

To evaluate the usability of an AI-enabled decision-support tool in a real operational setting, we conducted a mixed-methods study addressing these three
research questions:

\noindent \textbf{RQ1}:
What experiences and perceptions do analysts report when using an AI-enabled crime-linkage tool?

\noindent \textbf{RQ2}:
Do analysts attend to all the features \vn{underlying AI predictions when interacting with the tool?}

\noindent \textbf{RQ3}:
To what extent do analysts integrate their traditional \vn{analytical} methods to complement AI predictions?

Answering these questions required substantial preparation of real crime materials and the design of a mixed-methods evaluation strategy, including qualitative and quantitative methods. We adopted a four-pronged approach:
(i) direct behavioural observation of analysts interacting with the tool (addressing all RQs);
(ii) qualitative feedback to capture analysts' interpretations and sense-making strategies (addressing all RQs);
(iii) eye-tracking as an objective measure of analysts' interaction with AI predictions and their underlying features to address RQ2; 
(iv) mouse-tracking as an objective measure of analysts' interaction with the behavioural matrix to address RQ3; and
(v) established questionnaires as part of post-session surveys to assess analysts' perceptions of the usability and acceptability of the tool (addressing all RQs).

\subsection{Mixed methods usability study}

Our mixed-methods usability study involved participants interacting directly with the tool while we conducted real-time observation of their behaviour and collected eye-tracking and mouse-tracking data. We employed a repeated-measures (within-subjects) design, which means the same participants took part in different study elements. The study was designed to evaluate the tool through analysts interacting with its visualisation three times, each time with one of three different crime series. The crime series differed in their complexity and features that indicated if the crimes were linked or not. 



The study received ethical approval from the STEM Ethics Committee at the university of the researchers who conducted the study (University of Birmingham).
The SCAS analysts were sent an invitation email to take part in the study, together with an information sheet and a consent form. The information sheet stated that the study would take up to four hours and described evaluation steps.


\vn{For this study, it was not feasible to include all potential participants due to practical constraints. We had to accommodate both practical capacity constraints, such as the number of sessions that could realistically be conducted within a two-week period, and participant availability, ensuring that analysts were available to attend and that no participant completed more than one session per day.} Six participants (four female and two male, three analysts and three senior analysts) volunteered to take part.


Prior to the study, we conducted an online familiarisation workshop to explain how the tool worked, what each element represented, and how to interact with it. Participants were also provided with a written set of instructions to refer to afterward.


The usability study took place on a secure site where the participants worked, as they interacted with the tool using real crime data of a highly sensitive nature.
For each series, at least one crime linked to the index offence needed to appear within the top 20 of the ranked probability list. The analysts and researchers conducting the study were not told that there was a linked offence within the ranked list since in practice, this would also not be known to them. 

Although the underlying algorithms operate on tens of thousands of crimes—producing ranked lists that can span millions of rows—analysts reported that, in practice, they typically examine a subset of the highest-ranked crimes at any given time. Therefore, displaying only the top 20 crimes closely reflects their real-world workflow while keeping the task manageable.

During the study, the participants worked through the task of analysing data shown in the tool at their own pace and completion of the session typically lasted between 75-120 minutes. The first session was longer than the second for all participants as the index offence for the first session was more behaviourally rich than for other sessions.

In total, 16 experimental sessions were completed; four participants completed the experiment with the visualisation for all three cases and due to unforeseen circumstances two participants completed only two experimental sessions (with cases 1 and 2). Therefore, we had 16 sets of output in total and the participants completed the experiments in the same order.  

\noindent \textbf{Direct observation.}
Two researchers were present during the experimental sessions and took written notes on participants' behaviour and verbal comments when interacting with the tool. This approach was necessary because participants could be unaware of their real-time interactions, resulting in an incomplete understanding if relying solely on self-reports. Moreover, while eye- and mouse-tracking capture interaction patterns, they do not record participants' verbal comments or thought processes while performing a task. Direct observation therefore remained essential for developing a comprehensive understanding of users' engagement with the tool.

\noindent \textbf{Eye tracking.}
Eye-tracking data were recorded using an Eyelink 1000 SR research eye-tracker\footnote{\url{https://www.sr-research.com/eyelink-1000-plus/}} with the desktop mount and chinrest set-up, and using Weblink screen recording software~\footnote{\url{https://www.sr-research.com/weblink/}}. The eye-tracker recorded right-eye gaze location with a sampling rate of 1000 Hz during binocular viewing.

The tool was presented in full screen resolution on a 23.8 inches ThinkVision T24i-10 monitor with a refresh rate of 60 Hz and a screen resolution of 1920x1080 pixels. The viewing distance of the participant to the monitor was 1.75 times the width of the monitor screen (95 cm). The eye-tracker was set up to only record what the participants looked at on the monitor displaying the tool's output. Limiting the ranked list to 20 items ensured that the entire display remained on a single fixed screen, without the need for scrolling. Analysts could look off-screen if they wanted to interrogate another database that they commonly used in their analyses,
which was not recorded by the eye tracker.

\noindent \textbf{Mouse tracking.}
Tracking of mouse clicks was implemented as a temporary functionality in the tool for experimental purposes. The functionality recorded: (i) which case an analyst was viewing (the index offence), (ii) which cases they selected to compare with the index offence, (iii)
the duration for which the behavioural matrix pop-up remained open once activated, and (iv) how many times the behavioural matrix was activated.

Eye tracking and mouse tracking enabled a detailed analysis of user interaction patterns and provided insights that could not be captured through self-report alone.

\subsection{Post-session surveys}

After each session, participants completed three established questionnaires commonly used to capture users' perceptions while interacting with tools. We administered the surveys after every session rather than only at the end of the study because the tool might perform differently across crime series, and participants' perceptions could change as they became more familiar with using it.

\noindent \textbf{The Usefulness, Satisfaction, and Ease of Use (USE) questionnaire}~\cite{Lund2001Measuring}. The USE focuses on assessing the usability of systems. To reduce participant burden, participants rated 19 of the 30 original items about Usefulness, Ease of Use, Ease of Learning, and Satisfaction on a 7-point Likert scale from strongly disagree to strongly agree.

\noindent \textbf{The Evaluating User Experience in Information Visualization (UXIV) questionnaire}~\cite{victorelli2023Evaluating}. User experience (UX) influences the effectiveness of information visualisation (IV). UXIV is designed to evaluate user experience based on how they interact with data. One module of the UXIV is human-data interaction (HDI). A total of 13 items out of original 28 from the UXIV HDI module were selected by the research team based on their relevance to this evaluation. Items were measured on a 7-point Likert scale from strongly disagree to strongly agree. 

\noindent \textbf{Questionnaire for User Interface Satisfaction (QUIS)}~\cite{chin1988Development}. The QUIS is designed to measure user satisfaction with the human-computer interface. The questionnaire comprises 27 items across five sections: screen, terminology and system information, learning, system capabilities, and overall reactions to the software. The six items relating to the overall reactions to the software were included in this evaluation. These items measured reactions on a scale of 0-9 (with 0 denoting the most negative and 9 denoting the most positive response), with an additional ``not applicable'' option.

The participants were provided with a free text box in which they could outline any additional feedback they had about the visualisation (e.g., features they liked/disliked, aspects that aided their understanding, anything that was unclear). Free text responses were then analysed and presented as part of our findings.

\subsection{Data issue identified during the study}

During the study, it became clear that the data being displayed on the radar plot was not correct, which was raised by participants. Although we needed to discard the data and the results related to the radar plot from the paper as they could not be relied upon, our findings show that participants remained highly engaged with the data and all aspects of the tool's output.

\section{Findings}
\label{sec:results}

In this section, we present the findings of our mixed-methods usability study and address our research questions.  

\subsection{RQ1: Analysts' reported experiences and perceptions of the tool}

\begin{figure*}[t]
\centering
\includegraphics[width=\textwidth]{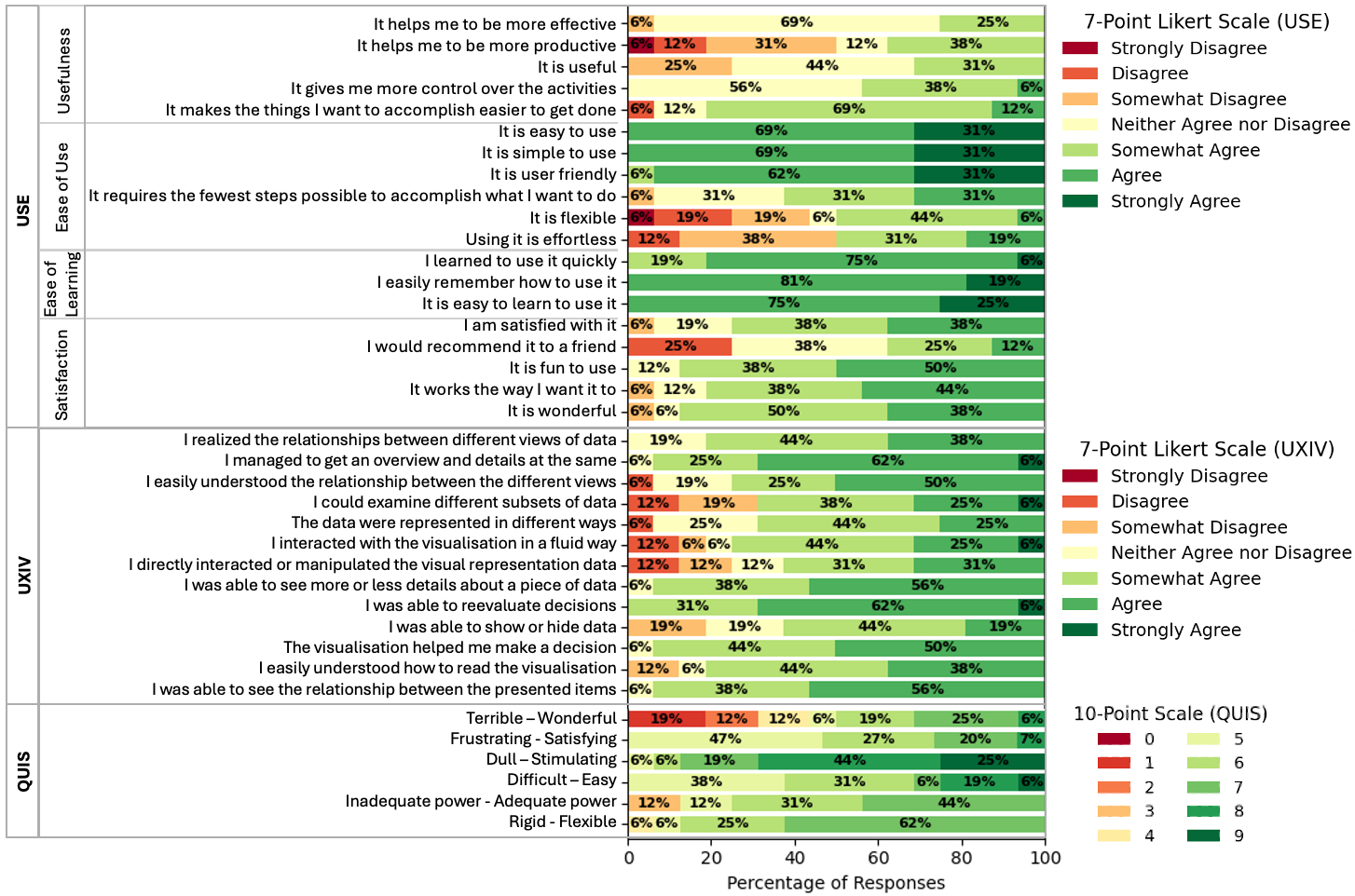}
\caption{Survey responses across 16 sessions combined, presented in a 7-point Likert scale from Strongly Disagree to Strongly Agree for USE and UXIV items, and from 0 (the most negative response) to 9 (the most positive response) for QUIS items.}
\label{figure:responsees}
\end{figure*}

To answer RQ1, we analysed the data collected through the direct observation of analyst behaviour during evaluation sessions, as well as their quantitative responses in post-session surveys.

\noindent \textbf{Direct observations.} According to the notes taken during the direct observation of participants, all participants engaged with the tool, and none appeared to lose interest as they progressed through the task. Interaction with the tool was unprompted and participants all used the tool throughout their sessions. They moved freely between the tool and other data sources that they routinely use in their analyses. This was encouraging to see because this was how we anticipated the tool would be used in a real-world setting.  

\noindent \textbf{Quantitative data from post-session surveys.} The analyses from the quantitative survey items are presented in Figure~\ref{figure:responsees}. It is apparent that the overall view of the tool was positive, while ease-of-use and ease-of-learning items had the highest ratings.
However, participants did not show full endorsement across all items, as reflected in the range of minimum and maximum scores for each statement, indicating variation within the sample of participants. This variation was particularly evident for items such as \textit{``It helps me to be more productive''} and \textit{``I interacted with the visualisation in a fluid way''}. This indicates that the tool needs further development to fully meet user needs, like to provide a more intuitive interaction and to make users more confident that the tool optimises their time spent on analytical tasks.

Additionally, we analysed USE responses for each session separately, which indicated that overall scores increased from session 1 to session 2, and then increased further in session 3. Because the tool did not change between sessions, this improvement likely reflects participants' growing familiarity and comfort with the tool.

\noindent
\\
\fbox{%
\begin{minipage}{0.46\textwidth}
\setlength{\parskip}{0.1em}
\setlength{\parindent}{0pt}
\vspace{1mm}
Results from the post-session survey and observational study indicate that participants engaged with the tool and generally perceived it positively. Ease-of-use and ease-of-learning items received the highest ratings. In contrast, perceived usefulness ratings were mixed, suggesting that additional iterations are needed \vn{prior to wider operational deployment}.
\vspace{1mm}
\end{minipage}%
}

\subsection{RQ2: Analyst attention to AI model features}


\begin{figure}[t]
\centering
\includegraphics[width=\columnwidth]{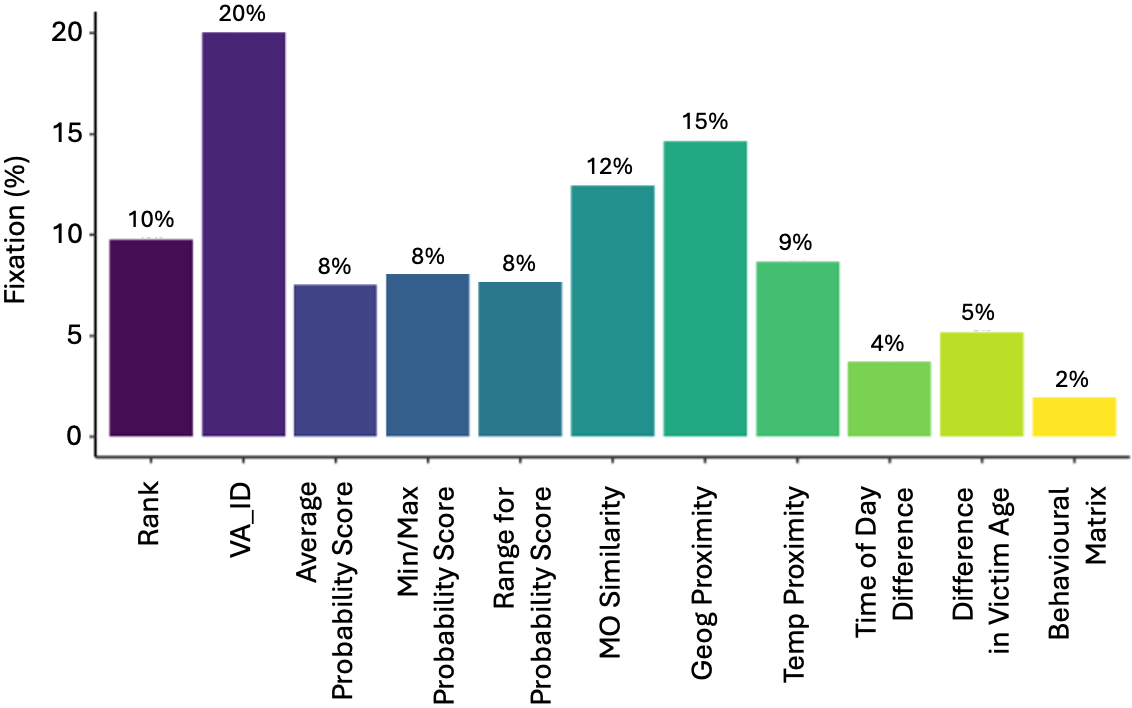}
\caption{Percentage of fixations for each element of the visualisation. \textit{Behavioural Matrix} represents the button on the main screen used to show the matrix.}
\label{figure:fixation}
\end{figure}

To answer RQ2, we performed the analysis of qualitative data collected in post-session surveys and observational study, as well as eye-tracking data collected during analysts' interaction with the tool.

\noindent \textbf{Qualitative data in post-session surveys and direct observations.} 
The written survey feedback and the verbal comments given during the observational study were analysed qualitatively. The results are presented as either positive feedback or recommendations for improvement. For confidentiality reasons (as required by our ethics approval), we do not included quotes, but we indicate if a finding comes from one or more participants, as shown by the corresponding \textit{n} value.

\noindent Several tool aspects were given positive feedback: 

\begin{itemize}
	\item The presentation of the different features that had contributed to the probability score in the ranked list (e.g., MO similarity or geographical proximity). Analysts valued having this information presented in one place and displayed side by side, rather than having to conduct separate searches to locate it, as is typically required in their traditional workflow. (\textit{n = 1})
  \item Use of colour intensity to draw attention to features that made a greater contribution to the probability score. A participant commented that this drew their attention to key information to which they needed to attend. (\textit{n = 1})
  \item The presentation of bar charts alongside the actual probability scores. A participant commented how this helped crimes that had a high probability of being linked to the index offence ``jump out''. (\textit{n = 1})
	\end{itemize}

\noindent Two improvements were suggested: 
\begin{itemize}
	\item Ability to grey-out or mark entries in the ranked list that they have already considered to allow analysts to book-mark progress down the ranked list. This was particularly important when analysts are switching between screens to analyse other data sources available to them. (\textit{n = 1})
	\item Removal of features in the ranked list that were believed to be irrelevant for a particular case when making decisions about crimes being linked or not.
  (\textit{n = 1})
\end{itemize}

The positive feedback shows that analysts appreciate aspects of the tool that display the features used by the AI model to generate its predictions. The suggested improvements indicate that analysts want the tool to better support focus and efficiency by allowing them to filter out crimes and features that are not relevant to their current decision-making.

\begin{figure*}[t]
\centering
\includegraphics[width=\textwidth]{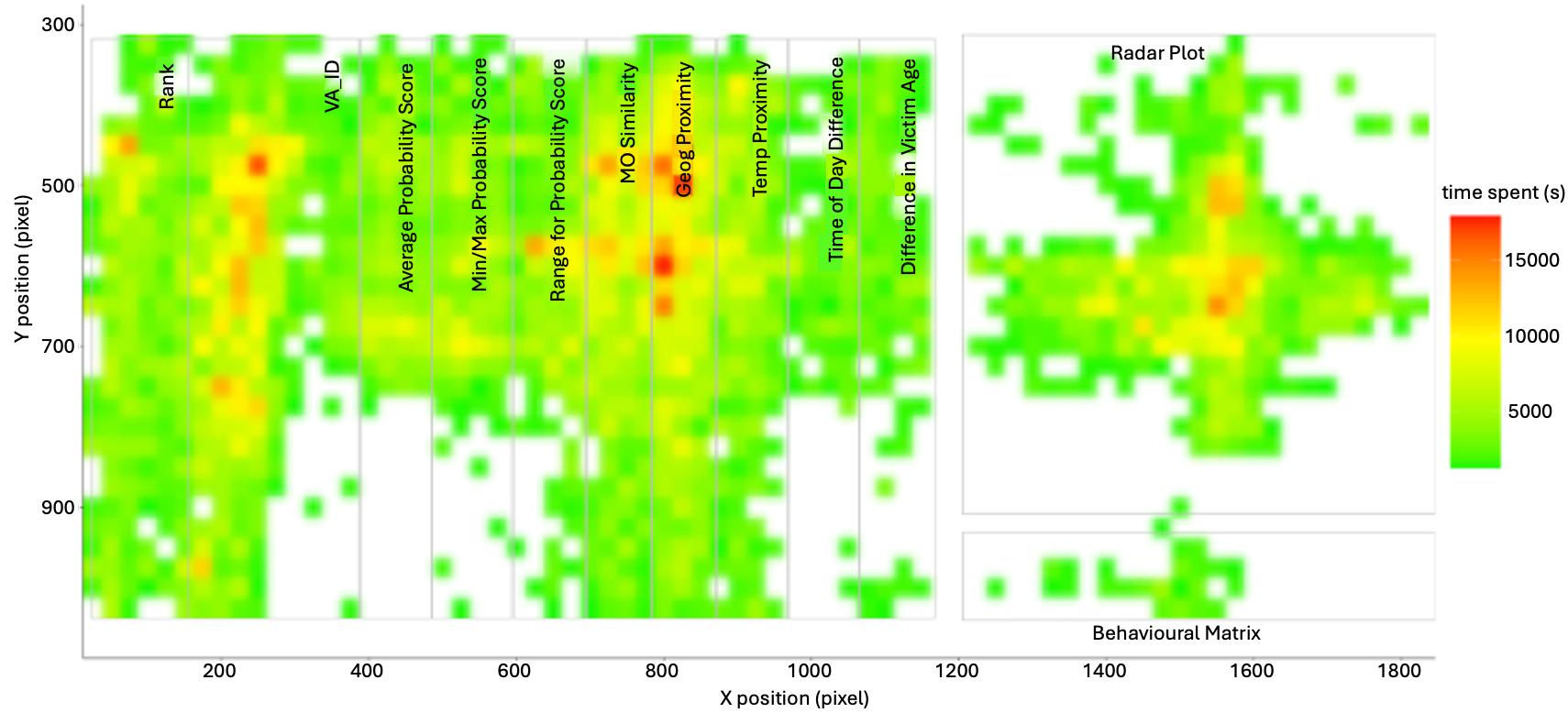}
\caption{Heat map for fixations on the tool's visualisation for all sessions and all participants combined. Behavioural matrix represents the button on the main screen used to show the matrix.}
\label{figure:heatmap}
\end{figure*}

\noindent \textbf{Eye-tracking findings.} We analysed eye-tracking data to determine the parts of the tool that received the greatest visual inspection by participants. These findings are shown as bar charts for proportion of fixations per area of the display in Figure~\ref{figure:fixation}, and as heat maps in Figure~\ref{figure:heatmap}.

Figure~\ref{figure:fixation} presents the percentage of total fixations by all participants and all cases combined for each element of the display. Those with a larger percentage therefore were fixated on more by participants than those with a smaller percentage. The figure indicates that all elements of the visualisation received some fixations. The element of the initial visualisation most fixated on by participants and across all cases was the VA\_ID, which is the unique reference number for a case allowing the analyst to retrieve further information about the crime from other data sources. They are therefore using the visualisation to determine cases with a high probability of being linked to the index offence from the visualisation, extracting this key information from the visualisation as well, and moving to another screen to retrieve additional information to inform their decision-making. This is exactly the use of the tool we were hoping to observe. The next most often fixated on element was geographical proximity, followed closely by MO similarity. These are two of the five features that contribute most towards the probability score for a pairing between a crime and the index offence and they are the features that analysts will be most familiar with. The other three features appear much less frequently in the published literature and the analysts are less interested in these. 


We also analysed data split by session, with each session corresponding to the analysis of one crime series. In general, a consistent pattern of fixation is seen across the three sessions, such as that the crime identifier (VA\_ID), MO similarity, and geographical proximity have the highest percentage of fixation in exactly that order for each session.

The heatmap for all cases and participants combined (see Figure~\ref{figure:heatmap}) indicates similar findings to those seen in the bar charts whereby hotter colours (red) are associated with more time spent looking at features such as VA\_ID, MO similarity, and geographical proximity. However, the heatmaps also indicate visualisation areas which were not looked at by analysts (white areas). From the heatmap, we can see that the analysts only spent time looking at some probability values and they stop looking at these at a certain point (indicated by white-coloured spaces). 

Interestingly, notable fixation happens in the geographical proximity and MO similarity columns towards the lower part of the display, which would draw analysts' attention even when reviewing the lower-ranked items. This suggests that these columns continue to be inspected even after the overall probability values begin to tail off.

In addition, we produced heat maps per session and per participant to determine if the findings were case-specific or participant-specific. The finding was reproduced across both these conditions. 

\noindent
\\
\fbox{%
\begin{minipage}{0.46\textwidth}
\setlength{\parskip}{0.1em}
\setlength{\parindent}{0pt}
\vspace{1mm}
Analysts responded positively to the availability of the AI model features presented as explanations, and suggested refinements to improve focus and efficiency during review. Eye-tracking confirmed that analysts attended to all explanatory features, with MO similarity and geographical proximity receiving the greatest attention, even for lower-ranked candidate links. As these features also contributed most strongly to the model's predictions, the findings suggest that the tool successfully exposes decision-relevant factors that align with analysts' validation practices, supporting transparency and informed use of AI outputs.
\vspace{1mm}
\end{minipage}%
}

\subsection{RQ3: Integration of traditional analytical methods.}


To answer RQ3, we analysed further qualitative data collected in post-session surveys and observational study, as well as mouse-tracking data collected during analysts' interaction with the visualisation.
Here we focus on the findings related to the behavioural matrix, as it represents a traditional
behaviour-based method used by analysts in their crime-linkage tasks.

\noindent \textbf{Qualitative data in post-session surveys and direct observations.} 
The analysis of written survey feedback and verbal comments provided during the observational study offered insights and suggestions for improving the visualisation of the behavioural matrix.

\noindent Positive feedback: 

\begin{itemize}
  \item The information provided in the behavioural matrix. Analysts reported using the information within the matrix to support their decision-making in the experimental sessions. (\textit{n = 1})
	\item Hover function in the behavioural matrix whereby hovering over a cell led to the variable name representing the column being displayed in landscape which assisted viewing. (\textit{n = 2})
\end{itemize}

\noindent Suggested improvements: 
\begin{itemize}
	\item Ability to hide columns for behaviours not of interest in the behavioural matrix. (\textit{n = 3})
	\item Re-order the sequence of variables (i.e., the order of columns) in the behavioural matrix to represent an alternative order with which the analysts were more familiar. (\textit{n = 1})
	\item Group columns in the behavioural matrix together when multiple columns represent a theme of offending behaviour.
  (\textit{n = 1})
\end{itemize}

Analysts provided positive feedback on the behavioural matrix and the information it presents. Suggested improvements focused on increasing interaction flexibility, which could further enhance participants' focus and efficiency when performing crime-linkage tasks.

\begin{table}
\centering
\caption{Summary of behavioural matrix interactions per session (\textit{S}) and participant (\textit{P}). 
\textit{\# Opens} indicates the number of times the matrix was opened, 
\textit{\# Selected cases} shows the average
number of cases selected for comparison with the index case, and
\textit{Avg} and \textit{Total} show the average and total time (minutes) the window remained open.}
\label{tab:behavioural_matrix_interaction}
\begin{tabular}{c c c c c c}
\toprule
\multirow{2}{*}{S} &
\multirow{2}{*}{P} &
\multirow{2}{*}{\# Opens} &
\multirow{2}{*}{\# Selected cases} &
\multicolumn{2}{c}{Time (min)} \\
\cmidrule(lr){5-6}
& & & & Avg & Total \\
\midrule\
\multirow{6}{*}{1} & P1 & 10 &1.00& 0.61 & 6.07  \\ 
                   & P2 & 19 & 1.11& 0.44 & 8.35  \\
                   & P3 & 29 & 1.07& 2.69 & 77.96  \\
                   & P4 & 4 & 1.25& \textbf{11.70} & 46.82  \\
                   & P5 & 9 & 1.00& 0.74 & 6.67  \\
                   & P6 & \textbf{38}& 1.50 & 3.38 & 128.61  \\
 
\midrule
\multirow{6}{*}{2} & P1 & 24 & 1.00& 0.69 & 16.60  \\
                   & P2 & 22 & 1.00& 0.34& 7.44  \\
                   & P3 & 22 & 1.00 & 2.71 & 59.71 \\
                   & P4 & \textbf{1} & 1.00& 0.31 & 0.31  \\
                   & P5 & 11 & 1.00& 0.76 & 8.41  \\
                   & P6 & 18 & 1.06& 3.98 & 71.62  \\

\midrule
\multirow{4}{*}{3} & P1 &29 &1.76 &0.74&21.33 \\
                   & P2 & 20  & 1.00& 0.19 & 3.72 \\
                   & P3 & 22  & 1.00& 4.06 & 89.22 \\
                   & P4 & \textbf{1}  & 1.00& \textbf{0.18} & 0.18  \\
\bottomrule
\end{tabular}
\end{table}

\noindent \textbf{Mouse-click findings.}
\vn{To further investigate analysts' interaction with the behavioural matrix, for each session and participant, we present (i) the number of times the matrix was opened (and closed), (ii) how many cases were selected to compare with the index offence on average, and (iii) the average and total duration for which the behavioural matrix pop-up remained open once activated (Table~\ref{tab:behavioural_matrix_interaction}). Note that due to unforeseen circumstances participants P5 and P6 completed only the first two sessions, resulting in a total of 16 sessions.}




The average number of times the analysts opened the behavioural matrix per session ranged from once to 38 times. \vn{In 12 out of 16 sessions, analysts opened the matrix at least ten times. Although this indicates that most of the time participants frequently opened the matrix, they usually performed a single comparison, as in the vast majority of cases (260 out of 279 matrix openings), participants selected only a single case for comparison.}

\vn{
The average time for which the matrix was open in a session ranged from 0.18 min (11 seconds) to 11.7 minutes. On average, the behavioural matrix was open for 2.52, 1.67 and 1.59 minutes per participant for sessions 1, 2 and 3, respectively. Participants had the matrix activated for a significantly shorter time in sessions 2 and session 3 in comparison to session 1 (34\% and 37\%, respectively). This decline in time can partly be explained by the index offence in session 1 being more behaviourally complex and partly by analysts' growing familiarity with the tool. The higher complexity of the first index offence likely required analysts to spend more time inspecting individual behaviours within the matrix.
However, no notable difference in behavioural complexity was observed between the index offences in sessions 2 and 3; yet we still see a decline in the time spent fixating on the matrix. This additional decrease likely reflects growing familiarity with the tool and needing less time viewing the matrix.}
%

\noindent
\\
\fbox{%
\begin{minipage}{0.46\textwidth}
\setlength{\parskip}{0.1em}
\setlength{\parindent}{0pt}
\vspace{1mm}
Analysts expressed positive feedback on the behavioural matrix and the non-AI behavioural information it provides, describing it as essential for validating candidate links. Suggested improvements concerned increasing interaction flexibility, which could further enhance analysts' focus and efficiency during crime-linkage analysis. Mouse-tracking data showed that analysts frequently verified AI predictions by cross-checking AI-ranked candidates against traditional behavioural analysis supported by the behavioural matrix, indicating that AI outputs are used as decision support rather than as a substitute for established practice. Overall, these findings highlight the importance of engineering AI-enabled crime-linkage tools that tightly integrate AI predictions, explanations, and non-AI behavioural evidence within the workflow to support verification, trust, and operational acceptance.
\vspace{1mm}
\end{minipage}%
}
\\

Our findings were shared with SCAS senior leadership and a list of refinements for the tool were agreed and are under development.

\section{\vn{Addressing ethical considerations}}
\label{section:ethics}

\vn{The use of AI in decision-making raises several ethical concerns, with bias, transparency, and accountability being recognised as the most crucial ones to address~\cite{Osasona2024EthicalAI}. To mitigate potential bias, the data used to train the AI models integrated into the crime-linkage tool were assessed for demographic representativeness in our prior work by Zhan et al.~\cite{zhan2026enhancingbinaryencodedcrime}, with no evidence of bias identified. Transparency is supported by presenting probability scores and non-AI behavioural data alongside AI-generated predictions. Additionally, the tool was explicitly designed as a decision-support system rather than an automated decision-making system. Its outputs are intended to assist analysts in identifying potential linkages, while final descisions remain the responsibility of human analysts. Accordingly, the use of AI in this context is not intended to replace human expertise, but to support analytical processes within appropriate organisational, legal, and ethical frameworks.}

\section{Threats to Validity}
\label{sec:threats}

\noindent \textbf{Construct validity.} Analysis of the written feedback revealed that participants did not always limit their comments to the usability of the tool. Several responses referred instead to aspects of the underlying algorithm or to desired extensions of the decision-support tool. Although such feedback is valuable for guiding the further development of the tool, it was out of the scope of this study and therefore excluded from the analysis.

\noindent \textbf{Internal validity.} Care was taken to ensure the analysts were not already acquainted with the three crime series in the evaluation, and the researchers administering the experiment and the analyst participants were blind to there being a linked crime in the output for each series. This design was intended to minimise potential bias arising from prior knowledge or expectations.

To address interpretive bias in analyses of survey responses, two authors of this paper independently analysed the data. All disagreements were resolved through discussion, and the findings were then reviewed and validated by the SCAS management.


\vn{Additional potential threats to internal validity include presenting tasks in the same order and differences in task complexity. In our study, overall questionnaire scores increased over time despite the tool remaining unchanged, suggesting that participants became more familiar and comfortable with the tool in each session, which is a potential manifestation of learning bias. Moreover, the first task was more complex than the others and the participants needed more time to complete it. Although task complexity should be considered when interpreting our findings, the tasks were deliberately designed to reflect the variety of complexity typically encountered by analysts in their work.}


\noindent \textbf{External validity.} Our usability study was conducted within one law enforcement team (SCAS), as the tool was developed specifically for this team. This context dependence may limit the generalisability of the findings to other settings or standalone uses of the tool. \vn{However, the finding of this study align with our findings from work with another team~\cite{alrajeh2026datadependentgoalmodelingmlenabled, Nowack2025UserCentredAI}, in particular about providing non-AI evidence alongside AI predictions.  
We believe that the findings of our study provide valuable insights into the evaluation processes of AI-enabled decision-support tools.}

\noindent \textbf{Conclusion validity.} Due to the issues with the algorithm that computed the data for the radar plot, it was not possible to present reliable findings regarding the radar plot component. However, resolving these issues provides a clear opportunity for future work to strengthen and more fully evaluate this aspect of the tool.

\section{Related Work}
\label{sec:related}

\noindent \textbf{AI-enabled decision support in law enforcement.}
\vn{Halford and Gibson~\cite{Halford2025UsingML} presented an ML-based crime linkage tool for linking residential burglaries and highlighted the importance of geographic and temporal features for the generated predictions. The authors, however, did not provide a user study, and thus provided limited insight into the tool's practical usability.}

Recent studies in the law enforcement domain have investigated which features are essential for AI-enabled systems to ensure successful adoption.
Dempsey et al.~\cite{Dempsey2023Exploring} reported mixed perceptions among police officers regarding AI, highlighting attributes such as fairness, accountability, transparency, and explainability as desirable for AI systems. In contrast, Herrewijnen et al.~\cite{Herrewijnen2024Requirements} found generally positive attitudes toward AI among police officers. Their exploratory study aimed to gain insights into perceptions of AI technologies overall, with a particular focus on explainable AI. Agudo et al.~\cite{Agudo2024} conducted user studies to examine how the timing of AI system support affects the accuracy of human decision-making.

Finally, Nowack et al.~\cite{Nowack2025UserCentredAI} presented an AI-enabled system and conducted a user study with crime analysts that showed that the system should satisfy requirements for scalability, accuracy, justification, trustworthiness and adaptability for its successful adoption. The study further underscored the importance of designing user-friendly human-AI interactions to support effective system use.

\noindent \textbf{Methods for evaluating expert decision-making.}
Eye tracking has been a common technique for user experience evaluation in the literature~\cite{Novak2024EyeTracking}.
Chang and Tsai~\cite{ChangTsai2022Visual} studied crime scene photograph inspection using eye tracking but sampled studies with police college students, rather than expert crime scene investigators. One study of experts' decision-making for blood spatter analysis also employed eye tracking but with a laboratory-generated (and therefore using real but mock) stimulus material outside the real-world setting \cite{Arthur2018EyeTracking}. Another study of forensic decision-making was conducted in a simulated crime scene~\cite{Baber2012Expertise}. However, the intended eye tracking element was abandoned due to fluctuations in ambient lighting, which highlights the challenge of bringing eye tracking methodology into a real-world setting.

Other authors conducted semi-structured interviews~\cite{Davies2018Practice,TonkinWeeks2021} and surveys~\cite{Burrell2011Preliminary} with analysts regarding their decision-making processes, and experimental studies using mock crime linkage tasks~\cite{Bennell2010Linkage,Canter1991Facet,Santtila2004Expertise}. While providing useful insights into human crime linkage decision-making, these approaches have limitations. For instance, studies utilising semi-structured interviews and surveys require participants to report on decision-making for historic cases. As such, data collected via these methods are prone to error, distortion and omission~\cite{Bradburn1987Answering,Elffers2010Misinformation}. Therefore, a mixed-methods evaluation is needed to provide a more comprehensive and reliable understanding of decision-making processes.


\noindent \textbf{Evaluating usability of AI-enabled systems.}
With the recent rapid advancement of AI technologies, researchers have increasingly focused on evaluating the usability of AI-based software. Some user studies were conducted with developers on AI-assisted programming systems. Vaithilingam et al.~\cite{Vaithilingam2023Towards} demonstrated how the iterative refinement of the system design guided by user feedback improved the usability and the acceptance of their tool. The findings presented by Liang et al.~\cite{Liang2024LargeScale} revealed that one of the key usability challenges was understanding which inputs contributed to the generated outputs. To address a similar issue in our system, we display feature-based explanations alongside generated predictions. 

Other studies have focused on the usability of AI-enabled decision-making systems. De Santana et al.~\cite{DeSantana2023Retrospective} found that visualisations combining predictions from multiple models can reduce usability due to their complexity. On the other hand, Kóvári~\cite{Kovari2024AIDecisionSupport} emphasised that providing explanations for AI models is essential for achieving user satisfaction. To address these issues, our tool presents relevant non-AI behavioural information alongside AI predictions, while keeping the number of visual elements limited to avoid overwhelming users.

\section{Conclusion}
\label{sec:conclusion}

In this paper, we presented a mixed-methods usability study evaluating an AI-enabled decision-support tool for crime linkage in the NCA. 
Our findings showed that analysts engaged actively with the tool and generally perceived it positively.
They responded favourably to the availability of model features presented as explanations for AI-generated predictions, which indicates that the tool successfully exposes decision-relevant factors that align with analysts' validation practices. 
In addition, analysts frequently verified AI predictions by cross-checking AI-ranked candidates against traditional behavioural analysis, indicating that AI outputs are used as decision support rather than as a substitute for established practice. 

Although participants suggested further improvements to increase interaction flexibility (which would enhance focus and efficiency during crime-linkage tasks), our findings overall highlighted the importance of engineering AI-enabled crime-linkage tools that tightly integrate AI predictions, explanations, and non-AI behavioural evidence within the workflow to support transparency, verification, trust, and operational acceptance.

\section*{Acknowledgement}
We acknowledge financial support from the National Crime Agency and the University of Birmingham EPSRC Impact Acceleration Award fund. We also thank the participants of our study for sharing their time, expertise and insights with us.%
%
%
\bibliographystyle{ACM-Reference-Format}
\bibliography{bib}
\end{document}